# Signature of Quantum Entanglement in $NH_4CuPO_4 \cdot H_2O$

**Tanmoy Chakraborty,** [1, a)] **Harkirat Singh,** [1] **Chiranjib Mitra.**[1, b)]

[1] Indian Institute of Science Education and Research (IISER) Kolkata, Mohanpur Campus, PO: BCKV CampusMain Office, Mohanpur - 741252, Nadia, West Bengal, India.

[a)] Electronic mail: tanmoy@iiserkol.ac.in.

[b)] Electronic mail: chiranjib@iiserkol.ac.in.

## Abstract

Entangled solid state systems have gained a great deal of attention due to their fruitful applications in modern quantum technologies. Herein, detection of entanglement content from experimental magnetic susceptibility and specific heat data is reported for $NH_4CuPO_4 \cdot H_2O$ in its solid state crystalline form. $NH_4CuPO_4 \cdot H_2O$ is a prototype of Heisenberg spin 1/2 dimer system. Temperature dependent magnetic susceptibility and specific data are fitted to an isolated dimer model and the exchange coupling constant is determined. Field dependent magnetization isotherms taken at different temperatures are plotted in a three dimensional plot. Subsequently, entanglement is detected both from susceptibility and specific heat through two different entanglement measures; entanglement witness and entanglement of formation. The temperature evolution of entanglement is studied and the critical temperature is determined up to which entanglement exists. Temperature dependent nature of entanglement extracted from susceptibility and specific heat shows good consistency with each other. Moreover, the field dependent entanglement is also investigated.

## I. INTRODUCTION:

Quantum magnets with low dimensional magnetic interactions have attracted considerable attention in various contexts from quantum computation to high energy physics [1-5]. In addition to studying magnetism in different spin models, a number of quite fascinating features of condensed matter physics, such as high temperature superconductivity [6], quantum Hall effects [7], heavy fermionic physics [8] etc. can be well studied on these systems. A variety of spin models have been proposed theoretically which have successfully explained experimental observations with high accuracy [9, 10]. Uniform spin chain, spin ladder, transverse spin Ising model etc. are some of the extensively investigated quantum spin systems where several aspects of quantum magnetism can be studied. In particular, spin ½ Heisenberg antiferromagnetic systems reflecting linear chain characteristics are potential candidates from the perspective of quantum information science [2]. One special case of such systems is dimerized spin ½ chain where interdimer interaction is negligible as compared to the intra-dimer interaction. Such systems closely approximate the two qubit system formulated theoretically [11, 12]. In the present work, such a physical bipartite system is studied where the tools of quantum information processing are applied to study quantum entanglement.

Quantum entanglement is an important resource in quantum computation and quantum information theory [13]. A great deal of research activities has been devoted to study entanglement both qualitatively and quantitatively in theoretical and experimental fronts [2, 14-16]. Entanglement is a curious phenomenon which is solely quantum in nature and does not have any classical analogue. A considerable number of research works are going on regarding entanglement characterization in condensed matter systems. Importantly, entangled states have been detected successfully in bulk materials in the thermodynamic limit [17-21]. Detection of entanglement in solid state crystalline form is a very necessary and important condition for the physical implementation of the proposed architecture of a feasible quantum computer [22]. Quantum correlation is a manifestation of collective behavior of interacting many-body quantum systems. Entanglement is a type of quantum correlation which naturally exists in physical systems. For instance, quantum spin systems provide an excellent playground for studying entanglement. The evolution of entanglement with different parameters, namely, temperature, magnetic field, anisotropy parameter etc. can be studied on these systems. As illustrations, entanglement content has been investigated in different spin systems like Heisenberg systems [2], 1-dimensional lattice models [22, 23], Anderson model [24] etc. Moreover, recently it has been observed that quantum entanglement has close connection with quantum phase transition [14, 25]. Entanglement plays a significant role at the quantum phase transition where quantum fluctuations appear in all length scales [26, 27]. Quantum correlations between the microscopic constituents of a solid state system can affect its macroscopic thermodynamic properties. As an example, for a magnetic system comprising spin ½ particles arranged in a lattice, magnetic susceptibility is capable to reveal spin entanglement between its constituents [28]. Thus a fruitful link has been established between quantum mechanics and thermodynamics and one can encapsulate information about entanglement by only carrying out basic thermal and magnetic measurements. There are evidences where thermodynamic properties have been used to detect entangled states [17, 18 and 29]. This is worth mentioning that one can easily obtain internal energy for a condensed matter system by carrying out specific heat measurements. However, in this scheme of detecting entanglement, compared to the magnetic susceptibility, internal energy is more useful due to its extendibility to non-magnetic systems. Numerous different propositions have been made for the estimation of entanglement from macroscopic observables [2]. Amongst them, entanglement of formation (EOF) is a well established method for the detection of bipartite entanglement. Moreover, another useful tool to determine the presence of entangled states is Bell's inequality test. Violation of Bell's inequality assures the existence of entanglement; although all entangled states are not bound to violate Bell's inequality [31, 32]. Thus, Bell's inequality violation can be used as an entanglement witness. In addition, another useful protocol for estimating entanglement from macroscopic thermodynamic variables is by constructing an observable called entanglement witness (EW) [33]. EW holds an empirical dependence with macroscopic observables. Thus, EW is able to provide the sufficient information whether a certain state is entangled or not [34]. The present work deals with the quantification of EW and EOF through both magnetic susceptibility and specific heat data.

The present compound under investigation is a spin-gapped system which can be well described by isolated Heisenberg spin ½ dimer model [35, 36]. Although the magnetic susceptibility data have been reproduced by isolated dimer model, experimental evidences indicate the existence of a very weak interdimer interaction. Previously reported crystal structure has efficiently established the correlation between the molecular structure and the magnetic behavior in $NH_4CuPO_4 \cdot H_2O$ [35, 36]. The crystal structure has revealed the fact that $NH_4CuPO_4 \cdot H_2O$ is a layered phosphate where two edge-sharing $CuO_5$ square pyramids pair up to form a spin ½ dimer which resides on the planner sheets of the layered structure. These $(CuO_5)_2$ dimers are connected in a crossed manner through $PO_4$ tetrahedras. The magnetic interactions in $NH_4CuPO_4 \cdot H_2O$ can be best described by the Heisenberg dimer Hamiltonian as given below

$$H = 2JS_1S_2 + B(S_1^z + S_2^z) \tag{1}$$

Here, $J$ is the exchange coupling constant and $B$ is the externally applied magnetic field. $S_1$, $S_2$, $S_1^z$ and $S_2^z$ are the site spins and the z-components of the spins at site 1 and 2 respectively.

By means of magnetic and thermal measurements, the signature of quantum entanglement has been detected in $NH_4CuPO_4 \cdot H_2O$ in the thermodynamic limit. Temperature dependent magnetic susceptibility, isothermal magnetization and temperature dependent specific heat data are collected for the present system. The magnetic and the specific heat data are analyzed within the framework of antiferromagnetic spin ½ Heisenberg dimer model. Using magnetic susceptibility and internal energy as EW, entanglement is detected and the critical temperature is determined up to which entanglement exists. Furthermore, the variation of entanglement with externally applied magnetic field is also captured. In the last section, EOF is quantified for $NH_4CuPO_4 \cdot H_2O$ from both magnetic susceptibility and specific heat.

## II. EXPERIMENTAL:

Copper (II) chloride dihydrate ($CuCl_2 \cdot H_2O$) and diammonium phosphate [$(NH_4)_2HPO_4$] of purest grade were obtained from Sigma Aldrich and used as starting reagents. Synthesis and crystallization were performed following the procedure described elsewhere [36]. An aqueous solution of $CuCl_2 \cdot H_2O$ (0.01M) was mixed with a saturated solution of $(NH_4)_2HPO_4$ (0.15M). Obtained mixture was preserved for three months in room temperature. Blue colored prismatic crystals were obtained which were separated from the solution by filtration, washed with diethyl ether, dried properly and carried to the next step for performing magnetic and thermal measurements.

Magnetic measurements were performed in a Magnetic Property Measurement System (MPMS) by Quantum Design, USA. Static magnetic susceptibility data were recorded in the temperature range of 2K to 100K. Subsequently, isothermal magnetization measurements as a function of magnetic field were carried out at different temperatures. Magnetic field was varied from 0T to 7T and the temperature was varied from 2K to 14K. Minimization of the trapped magnetic field was performed before starting each measurement. Standard relaxation method was employed to carry out the specific heat measurements. In relaxation method, a constant heating pulse is applied to an adiabatically insulated thin piece of crystal until the temperature of the sample reaches a steady state value. Subsequently, as the heat supply stops, the temperature of the sample starts decaying. Thus, by measuring the time constant of the decay one can measure the specific heat by following simple mathematical relationships [37]. The measurements were performed in a cryogen free magnet manufactured by Cryogenic Limited, UK. In absence of external magnetic field, specific heat data were collected in the temperature range of 2K to 8K.

### III. RESULTS AND DISCUSSIONS:

Fig. 1 displays temperature dependent magnetic susceptibility ($\chi$) over the temperature range of 2K to 100K. The most interesting feature in the susceptibility curve can be observed is a rounded maxima around T=6.5K followed by a gradual decrease with further enhancement in temperature. The aforementioned characteristic in $\chi$ vs. T curve is a sign of having antiferromagnetic correlations in the system which has been confirmed by fitting the isolated Heisenberg dimer model to the experimental data. By means of rigorous numerical simulations, Johnston *et al.* calculated temperature dependent magnetic susceptibility for antiferromagnetic

S=1/2 alternating-exchange Heisenberg chain over the entire range of the alternation parameter α ($0 \leq \alpha \leq 1$, where $\alpha = 0$ represents a dimerized chain and $\alpha = 1$ represents the case of a uniform chain) [9]. The general expression, formulated by them, was capable to fit the numerical data over the entire range of $\alpha$ ($0 \leq \alpha \leq 1$) and over a wide range of the reduced temperature *t* with an excellent accuracy and reasonable values of the parameters. In order to analyze the χ vs. T data for $NH_4CuPO_4 \cdot H_2O$, we have considered the dimerized chain case of expression where the exchange coupling constant $J$ and the Landé g-factor $g$ were used as fitting parameter in the fitting routine. The expression reads as,

$$\chi \approx \frac{Ng^2\mu_B^2}{4K_BT} \frac{\sum_{n=1}^{5} N_n/t^n}{\sum_{m=1}^{6} D_m/t^m} \tag{2}$$

With $N_1$=0.6342798982, $N_2$=0.1877696166, $N_3$=0.0336036173, $N_4$=0.003861106893, $N_5$=0.0002733142974, $D_1$=-0.1157201018, $D_2$=0.08705969295, $D_3$=0.005631366688, $D_4$=0.001040886574, $D_5$=0.00006832857434, $t = K_BT/2J$ and N= Avogadro's number. The fit yielded $g$ =2.11 and $J$ =5K which are supported by the previously reported values [35]. The theoretical fit (solid curve) and the experimental data (circles) are shown in the Fig. 1. It must be emphasized that the theory and the experiment are in good consistency with each other. In order to capture the dependence of entanglement on magnetic field, magnetization isotherms are collected at different temperatures. As entanglement is a low temperature phenomena and the critical temperature (the temperature up to which entanglement persists) has a close relation with the antiferromagnetic ordering temperature [28], the measurements are mainly constrained in the regime where antiferromagnetic correlations survive significantly. Since the maxima in the susceptibility curve appears around 6.5K, the magnetization isotherms are taken from 2K to 10K and the magnetic field is varied from 0 to 7T. Using these magnetization isotherms, one surface plot is yielded where magnetization is shown along the vertical axis. The temperature and the magnetic field are varied along the two horizontal axes. The surface plot (Fig. 2) is capable to describe the distinct feature of magnetization when temperature and magnetic field both are varied.

Experimental specific heat curve (in absence of field) in the temperature range of 2K to 8K is shown in Fig. 3. The most pronounced characteristic can be observed in the specific heat curve is a rounded peak at temperature $T_{max}$= 3.5K with a subsequent decrease upon further increasing the temperature. However, at higher temperature regime, an up rise can be observed. A possible explanation for this behavior of the specific heat is described as follows. In general, the specific heat for magnetic compounds can be described as an additive effect of three interplaying terms.

$$C(T) = \gamma T + \beta T^3 + C_m(T) \tag{3}$$

The first term in Eq. (3) represents the electronic specific heat which varies linearly with temperature. Its contribution to the total specific heat is determined by the Somerfield coefficient $\gamma$. $NH_4CuPO_4 \cdot H_2O$ being an insulating compound, electronic specific heat does not have any contribution in this case. The coefficient $\beta$ is responsible for the lattice contribution and the third term represents the magnetic component. The lattice contribution influences the total specific heat negligibly small at lower temperature [38]. Hence, the magnetic specific heat appears to be dominating over the lattice part in low temperature regime. The appearance of the rounded maxima in the specific heat curve at 3.5K is most likely due to the Schottky effect which reflects the characteristic of a two level system [38]. When the temperature is sufficiently low, the thermal energy is unable to excite the system to higher energy levels which results less specific heat. However, with increase in temperature, as the probability of excitation increases, the specific heat also increases. The probability of excitation reaches its maximum value when the thermal energy becomes of the order of energy gap of the system owing to the broad maxima in the specific heat curve. At higher temperature, energy levels become equally populated and no differential change in the internal energy occurs. Consequently, the specific heat starts decreasing gradually. However, upon further increase in temperature, an upturn can be observed in the specific heat curve which happens due to the dominating role of the lattice part over the magnetic contribution. This scenario has been analyzed within the framework of Heisenberg dimer model taking into account the lattice contribution. Now, the molar magnetic specific heat for isolated Heisenberg dimer can be expressed as [39]

$$C_m(T) = 12R \left(\frac{J}{K_B T}\right)^2 \frac{e^{\frac{2J}{K_B T}}}{1 + 3e^{\frac{2J}{K_B T}}} \qquad (4)$$

The magnetic part of the specific heat data was extracted by subtracting the lattice contribution from the total specific heat. Eq. (4) was fitted to the magnetic part of the specific heat data where $J$ was allowed to vary as free parameter. The fit generated J=5K (which supports well our previous analysis on magnetic data). The best consistency was found for $\beta = 0.00022\ K^{-3}$ which is close to its reported value [36]. It is worth noting that an excellent match is found between the theoretical fit and the experimental data (Fig. 3). With the known value of $\beta$, one can easily estimate the Debye temperature $\theta_D$ using the following relation [38]. We obtained $\theta_D = 101.9$ K.

$$\theta_D = \left(\frac{12\pi^4 N K_B}{5\beta}\right)^{\frac{1}{3}} \qquad (5)$$

Next, we will examine the existence of entanglement in $NH_4CuPO_4 \cdot H_2O$ through experimental susceptibility and internal energy data. We have aimed to study entanglement in the antiferromagnetic spin 1/2 compound $NH_4CuPO_4 \cdot H_2O$ for the following reasons. For an antiferromagnetic system, the order parameter (which is the staggered magnetization) does not commute with the Hamiltonian leading to spin fluctuations in the system [18]. Consequently, the antiferromagnetic ground state becomes maximally entangled which also contributes to non-zero entanglement even at finite temperatures. Thus, one can experimentally capture the existence of entanglement in the thermal states of the system. In the present case, we aim to investigate how entanglement is influenced by two external parameters; temperature and magnetic field. Favorably, $NH_4CuPO_4 \cdot H_2O$ has an exchange coupling strength of 10K which enables us to easily access the temperature and magnetic field range where the significant entanglement features of the system can be captured. Moreover, $NH_4CuPO_4 \cdot H_2O$ being a prototype of spin ½ dimer model, theoretical formulations applicable for general bipartite case [11, 30] can be efficiently tested on it. Detection of entanglement via EW has been described as a powerful entanglement measuring protocol by Herodekki *et al.* [33]. Wiesniak *et al.* have proposed that magnetic susceptibility can be used as an EW. They have established the applicability of EW on a wide range of magnetic systems [28]. Entanglement Witnesses are thermodynamic observables which are capable to capture the entanglement content present in a system. An observable W corresponding to a state ρ can be used as an EW if Tr (ρW) > 0, when ρ is an entangled state. When Tr (ρW) <0, ρ may or may not be entangled [34]. Thus, EW supplies a necessary and sufficient condition for entanglement detection. However, violation of that condition does not confirm separability. To estimate the pair-wise entanglement in the dimerized Heisenberg chain compound NH$_4$CuPO$_4 \cdot$H$_2$O, we introduce the mathematical expression of EW which holds a functional dependence on magnetic susceptibility [Eq. (6)]. Note that the applicability of the aforementioned witness is more general, i.e. entanglement can be detected using this method for a large class of spin systems [28].

$$EW = 1 - \left( \frac{6K_B T \chi}{(g\mu_B)^2 N} \right) \tag{6}$$

Here the symbols have their usual meanings; $T$ is the temperature, $K_B$ is the Boltzmann constant, $g$ is the Lande g factor and $\mu_B$ is the Bohr magneton. In the above equation, EW has been expressed for the isotropic case where the components of the magnetic susceptibility along X, Y and Z directions are equal to each other. Based on the EW criterion discussed so far, it is possible to impose a bound in the $\chi$ vs. $T$ graph which is capable to separate out the entangled region from the separable one. The dotted red line in the inset of Fig. 4 represents the bound. The existence of entanglement is determined by the inequality $\left( \frac{6K_B T \chi}{(g\mu_B)^2 N} \right) < 1$ which is represented by the left-hand side of the curve. However, the right-hand side of the curve, which is governed by the condition $\left( \frac{6K_B T \chi}{(g\mu_B)^2 N} \right) > 1$, does not confirm separability. Based on this formulation, one can determine the entanglement critical temperature below which the system remains entangled. By principle, entanglement can only have positive values. Hence, negative values of EW are assumed to be zero. One can clearly see from the plot that the bound intersects the susceptibility curve at 7.6K. This indicates that the entanglement critical temperature for $NH_4CuPO_4 \cdot H_2O$ is 7.6K. To gain quantitative insight into how EW evolves with temperature, we have extracted EW from the susceptibility data using Eq. (6). Fig. 4 depicts the explicit variation of quantified EW as a function of temperature. EW vanishes at 7.6K which is consistent with the previous analysis. Arnesen *et al.* have theoretically investigated pair-wise entanglement as a function of magnetic field and temperature considering Heisenberg spin dimer wherein the entanglement appeared to be influenced by temperature induced magnons [11]. As the temperature is increased, the proportion of separable triplet state increases. Thus, the relative contribution of entangled states reduces in the statistical mixture of entangled and separable states. Consequently, it so happens, that for zero field, entanglement attains its maximum value where temperature is minimum and decreases afterwards with increasing temperature which supports our previous experimental results. On the other hand, they also investigated the evolution of entanglement upon application of external magnetic field. The typical nature of decreasing entanglement is noticed when the magnetic field is increased. Increase in magnetic field also increases the contribution from the separable triplet state at the expense of entangled states, thereby reducing entanglement. Fig. 5 exhibits the theoretical plot where the variation of entanglement has been captured with field and temperature. It must be mentioned that Eq. (6) is only valid when zero field susceptibility is considered. In presence of applied magnetic field, the singlet state no longer remains the lowest energy state as the applied magnetic field causes elementary excitations and changes the energies of the eigenstates. Therefore, Eq. (6) takes the form [40]

$$EW = 1 - (\frac{2M_z}{g\mu_B N} + K_B T \frac{2\chi}{(g\mu_B)^2 N}) \quad (7)$$

To compare the theory with the experimental results, we have used the isothermal magnetization datasets in Eq. (7) to create a 3D plot (shown in Fig. 6) where entanglement is depicted as a function of both temperature and magnetic field. It is worth mentioning that a striking similarity is found between the theoretical and experimental surface plots.

In a recent report it has been demonstrated that internal energy can serve as an entanglement witness in the thermodynamic limit when the system remains in the thermal equilibrium [29]. Weisniak *et al.* have detected entanglement for the thermal states of transverse spin Ising model by minimizing the variance of the Hamiltonian over all separable states [29]. They have established the fact that for macroscopic bodies, entanglement content can be detected both from the internal energy and specific heat considering certain Hamiltonians. Since the specific heat is a well established experimentally measureable quantity and the internal energy can be easily evaluated from specific heat by using simple mathematical formula, we have used the internal energy as an EW in the present case. The relation between specific heat and internal energy can be written as

$$U(T) = U_0 + \int_0^T C(T) dT \qquad (8)$$

With $U_0$ being the ground state energy. By means of numerical integration on the magnetic part of the specific heat data (extrapolated down to 0K), we obtained the internal energy dataset as a function of temperature. The ground state energy $U_0$ was estimated theoretically and incorporated in the integration. For $NH_4CuPO_4 \cdot H_2O$, $U_0 = -3J/2 = -7.5K$ (considering the Hamiltonian for a Heisenberg dimer). Both the theoretical and the experimental energies are scaled in the unit of Kelvin. In view of our target to examine the presence of entanglement from internal energy, we now introduce the entity called concurrence which has been established as a good measure of entanglement. Concurrence is described by the following equation [12, 30].

$$C = \frac{1}{2}\max[0, \frac{2|U|}{NJ} - 1] \quad (9)$$

The above equation has been derived considering the pair-wise entanglement between two spin ½ particles. The relation described in Eq. (9) is of great importance in terms of setting up a connection between internal energy and entanglement. This relation enables us to estimate entanglement quantitatively in a physical system. Positivity of $C$ is governed by the condition $(2|U|/NJ) \geq 1$. Physically, the statement implies that entanglement can exist in a given system if and only if the above inequality holds. Therefore, using this condition we have introduced a bound in the internal energy curve. In the upper region of the bound, the system remains entangled. However, the lower region demarked by the condition $(2|U|/NJ) \leq 1$, does not confirm separability. The intersection of the bound and the internal energy curve is associated with the critical temperature up to which entanglement exists. The critical temperature is found to be 8.1K in this case. Moreover, the quantified values of EW using Eq. (6) have been plotted with temperature in Fig. 7. It can be clearly observed that entanglement shows a gradual decrease with temperature and comes down to zero at T=8.1K which is the critical temperature mentioned earlier. Thus, one can conclude that quite similar feature has been observed in the quantitative nature of entanglement extracted from heat capacity and magnetic susceptibility. The critical temperatures determined from both the analysis are also quite close to each other.

Entanglement of Formation (EOF) is widely accepted as a mathematically susceptible measure of entanglement between two spin ½ particles. The notion of EOF, as was first proposed by Wootters [30], tells that EOF estimates the non-local resources required to create a given entangled state. Here we use EOF to detect bipartite entanglement in $NH_4CuPO_4 \cdot H_2O$ single crystals. Estimating EOF in real physical systems is a highly challenging and difficult task. Nevertheless, for certain cases, namely, for the simple model of two qubit system, one can compute EOF following the protocol prescribed by Wootters. For a system in equilibrium with a thermal reservoir, the density matrix is defined as $\rho = \exp(-\beta H)/Z$. Here, $\beta = 1/K_B T$, $Z$ is the partition function and H is the two-spin Heisenberg Hamiltonian defined in Eq. (1). Presently, we restrict ourselves to the case B=0. Therefore, considering the standard basis of a two qubit system {|00>, |01>, |10>, |11>}, it is straightforward to obtain the density matrix $\rho$ and the spin-reversed density matrix defined by $\tilde{\rho} = (\sigma_y \otimes \sigma_y) \rho^T (\sigma_y \otimes \sigma_y)$, where $\sigma_y$ is the matrix $\begin{pmatrix} 0 & -i \\ i & 0 \end{pmatrix}$ and $\rho^T$ represents the transpose of the matrix $\rho$. Now, if R is defined as the product of the density matrix and the spin-reversed density matrix, $\lambda_1, \lambda_2, \lambda_3, \lambda_4$ be the square roots of the eigenvalues of R, then the quantity called concurrence is given by $C = \max\{\lambda_1 - \lambda_2 - \lambda_3 - \lambda_4, 0\}$ [11, 30]. An analytical relation has been established between EOF and $C$ and it has been shown that EOF increases monotonically with increase of $C$. Thus, EOF as a function of $C$ can be written as [11, 30]

$$EOF = -\frac{1+\sqrt{(1-C^2)}}{2} \log_2 \frac{1+\sqrt{(1-C^2)}}{2} - \frac{1-\sqrt{(1-C^2)}}{2} \log_2 \frac{1-\sqrt{(1-C^2)}}{2} \qquad (10)$$

The minimum value EOF can take is '0' which is associated with the separable state whereas EOF=1 indicates maximal entanglement. In general, for interacting spin ½ systems, the two site spin-spin correlation function can be defined as $G_{12} = <\sigma_1 \sigma_2>$. For the case of isotropic

Heisenberg Hamiltonian, which has a global SU(2) symmetry, one can readily check that $<\sigma_1^z\sigma_2^z>=<\sigma_1^y\sigma_2^y>=<\sigma_1^x\sigma_2^x>=G$. Now, the concurrence $C$ for isotropic Heisenberg model is empirically connected with the correlation function G through the following relation [16]

$$C = \frac{1}{2}\max\{0, 2|G| - G - 1\} \tag{11}$$

The magnetic susceptibility $\chi$ can be expressed in terms of the correlation function $G$ [41, 42] as given by $\chi = (N_A g^2 \mu_B^2 / K_B T)(1+G)$. Endowed with the abovementioned analytical forms of EOF, C and G, it is straightforward to calculate EOF from experimental magnetic susceptibility data.

Based on the criteria discussed so far, we have estimated EOF at finite temperature for the present system and captured its variation with temperature. Spin-spin correlation function $G$ was evaluated from the experimental susceptibility data using the relation $\chi = (N_A g^2 \mu_B^2 / K_B T)(1+G)$. The numerical values of G were used in Eqs. (11) and (10) which eventually enabled us to calculate EOF as a function of temperature. The EOF vs. T plot is shown in Fig. 8. At lowest temperature (2K), EOF attains its maximum value of 0.7 which subsequently decays off as temperature increases and becomes zero at higher temperature (~ 16K). This feature of pair-wise entanglement is supported by the observations made by M. C. Arnesen *et al.* [11]. Moreover, in addition to the magnetic susceptibility, internal energy was also employed to quantify EOF. The mathematical relation $U = -(3RJ/2K_B)G$ directly links internal energy $U$ to the spin-spin correlation function $G$ [41, 42]. Hence, calorimetric measurements enable us get the correlation function G which was further used in Eqs. (11) and (10) to determine EOF. Plot of EOF with T is shown in Fig. 9 where the temperature is varied from 2 to 10K. The circles represent the quantified values of EOF from experimental data and the solid line represents EOF calculated using the theoretical dimer model with same model parameters. It is evident from Fig. 9 that the extrapolation of the experimental data indicates that the critical temperature up to which entanglement can be detected in the present system is close to 16K. Hence, one can conclude that the nature of EOF in both the cases (quantified from U and $\chi$) are in excellent agreement with one another.

## IV. CONCLUSION:

Our discussions in the present paper are all about exploring entanglement, the most appealing quantum correlation, in a solid state bulk body. In brief, we have investigated experimental evidences of entanglement in a spin ½ antiferromagnet in the thermodynamic limit using entanglement detection protocols where macroscopic thermodynamic entities, namely, magnetic susceptibility and specific heat have been used to detect entanglement. Magnetic susceptibility and specific heat measurements are performed on the single crystals of $NH_4CuPO_4 \cdot H_2O$. In order to correlate the observed magnetic behavior with dimerized Heisenberg chain model, we have fitted the experimental susceptibility and specific heat data to the empirical expressions for spin ½ dimer model. We obtained J=5K which is consistent in both the analyses. Subsequently, well established theoretical formulations are used to make a quantitative estimate of the entanglement content present in the system. Firstly, entanglement content has been quantified from magnetic susceptibility and later verified through specific heat. Separable bounds are imposed on the susceptibility and internal energy curves to determine the critical temperature up to which entanglement persists. The temperature evolutions of EW (estimated from susceptibility) and concurrence (quantified through specific heat) have been represented graphically. A striking match of the critical temperatures is found between these two cases. Moreover, isothermally measured field dependent magnetization curves are used to capture the field variation of entanglement. Estimated entanglement is plotted in a surface plot with temperature and field along the two horizontal axes. In the last section, entanglement content in $NH_4CuPO_4 \cdot H_2O$ is investigated through EOF, another promising tool for entanglement detection. Herein, both magnetic and calorimetric data are used to quantify EOF. Temperature dependent behaviors of EOF are investigated in both the cases. An excellent match is found between these two results.

Having entanglement both in optical and solid state systems can have potential applications in quantum technologies. However, the possibility of having quantum mechanically entangled spins in a solid state crystalline material has an advantage over the optical systems as the crystals can be efficiently integrated with existing Si based technology or other quantum devices [43, 44]. In addition, spin systems can have fruitful applications from the perspective of quantum communications. Bose has described that an entangled spin system can be used as an appropriate channel for transmitting a quantum state over a short distance [45]. It has been suggested that the above scheme can be efficiently implemented for Heisenberg spin 1/2 compounds with nearest neighboring interaction. In this case, a quantum state can be transferred with an improved fidelity than the classical one [45]. Successful implementation of the above protocol can play a significant role in designing a feasible quantum computer.

**ACKNOWLEDGMENTS:**

The authors would like to thank the Ministry of Human Re-source Development (MHRD), Government of India, for funding. The authors also would like thank to Dr. Swadhin Mandal for allowing us to use his lab facilities for the synthesis of the system.**REFERENCES:**

[1] Y. Nishida, Y. Kato and C. D. Batista, Nat. Phys. **9**, 93 (2013).

[2] L. Amico, R. Fazio, A. Osterloh and V. Vedral, Rev. Mod. Phys. **80,** 517 (2008).

[3] T. Rööm, D. Hüvonen and U. Nagel, Phys. Rev. B **69**, 144410 (2004).

[4] F. Heidrich-Meisner, A. Honecker, D.C. Cabra and W. Brenig, Phys. Rev. B **66,** 140406(R) (2002).

[5] P. Lemmens, G. Guntherodt and C. Gros, Phys. Rep. **375**, 1 (2003).

[6] P. A Lee, Rep. Prog. Phys. **71,** 012501 (2008).

[7] G.-W. Chern, A. Rahmani, I. Martin and C. D. Batista, arXiv:1212.3617v1 (2012).

[8] A. Theumann and B. Coqblin, Phys. Rev. B **69**, 214418 (2004).

[9] D. C. Johnston, R. K. Kremer, M. Troyer, X. Wang, A. Klumper, S. L. Budko, A. F. Panchula, and P. C. Canfield, Phys. Rev. B **61**, 9558 (2000).

[10] D. C. Johnston, R. J. McQueeney, B. Lake, A. Honecker, M. E. Zhitomirsky, R. Nath, Y. Furukawa, V. P. Antropov and Y. Singh Phys. Rev. B **84**, 094445 (2011).

[11] M.C. Arnesen, S. Bose and V. Vedral, Phys. Rev. Lett. **87**, 017901 (2001).19

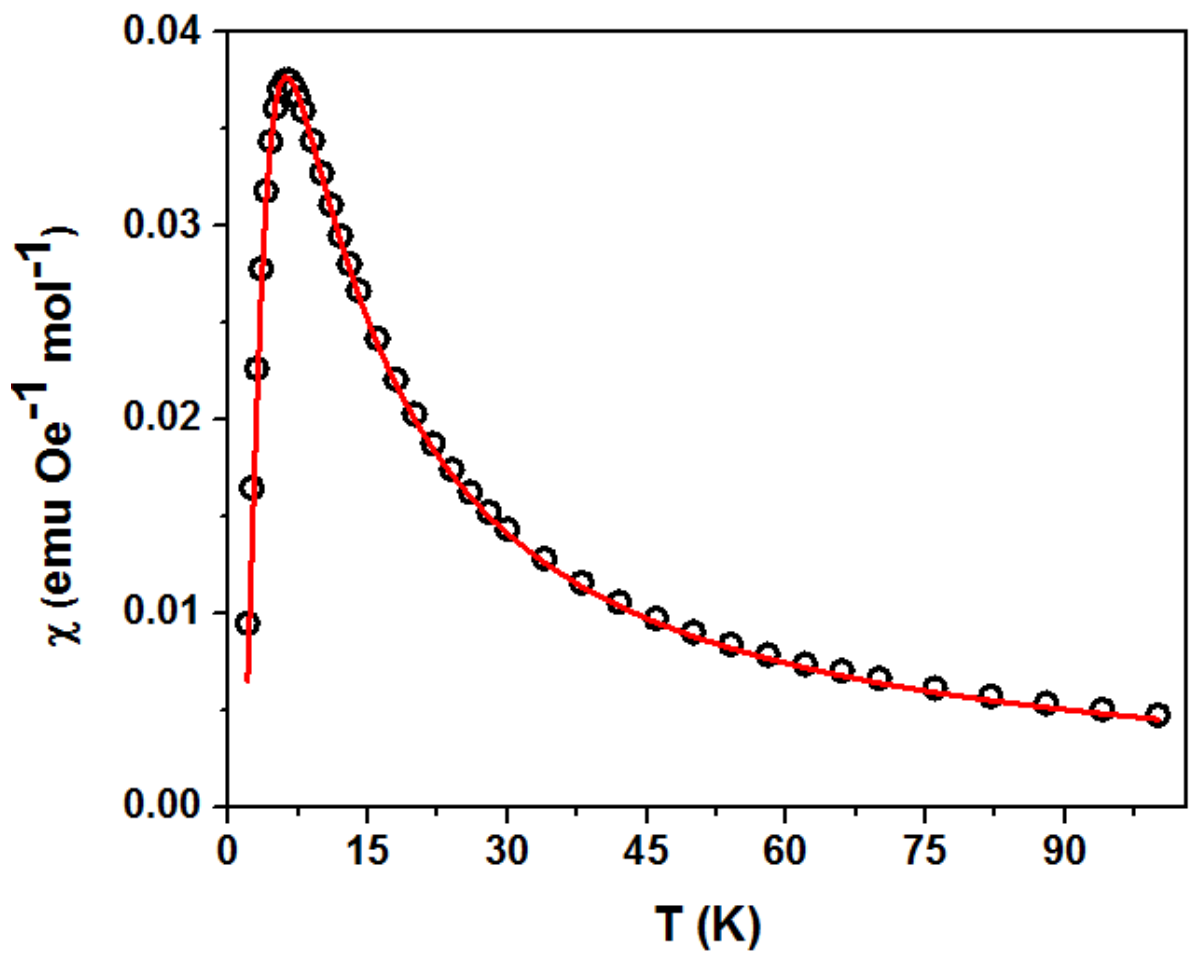

FIG. 1. Temperature dependent magnetic susceptibility for $NH_4CuPO_4 \cdot H_2O$. The experimental data are shown by the circles and the solid curve is the theoretical fit based on Eq. (2).

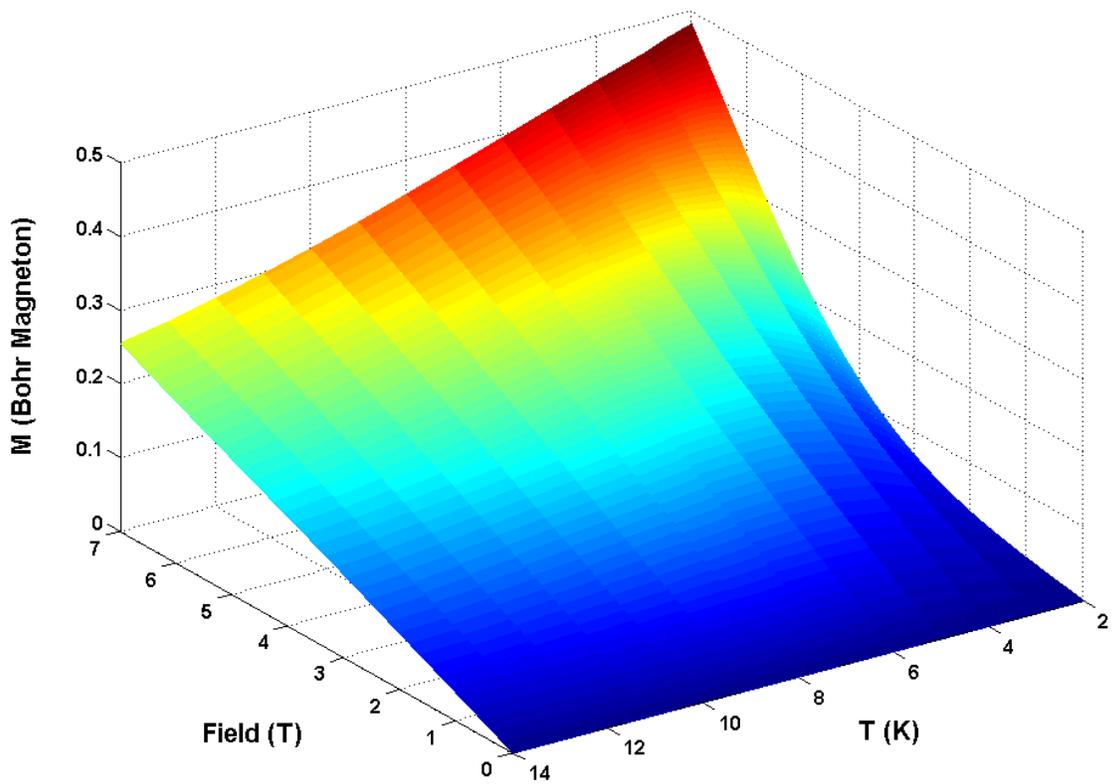

FIG. 2. Surface plot depicting the variation of magnetization with magnetic field and temperature for $NH_4CuPO_4 \cdot H_2O$.

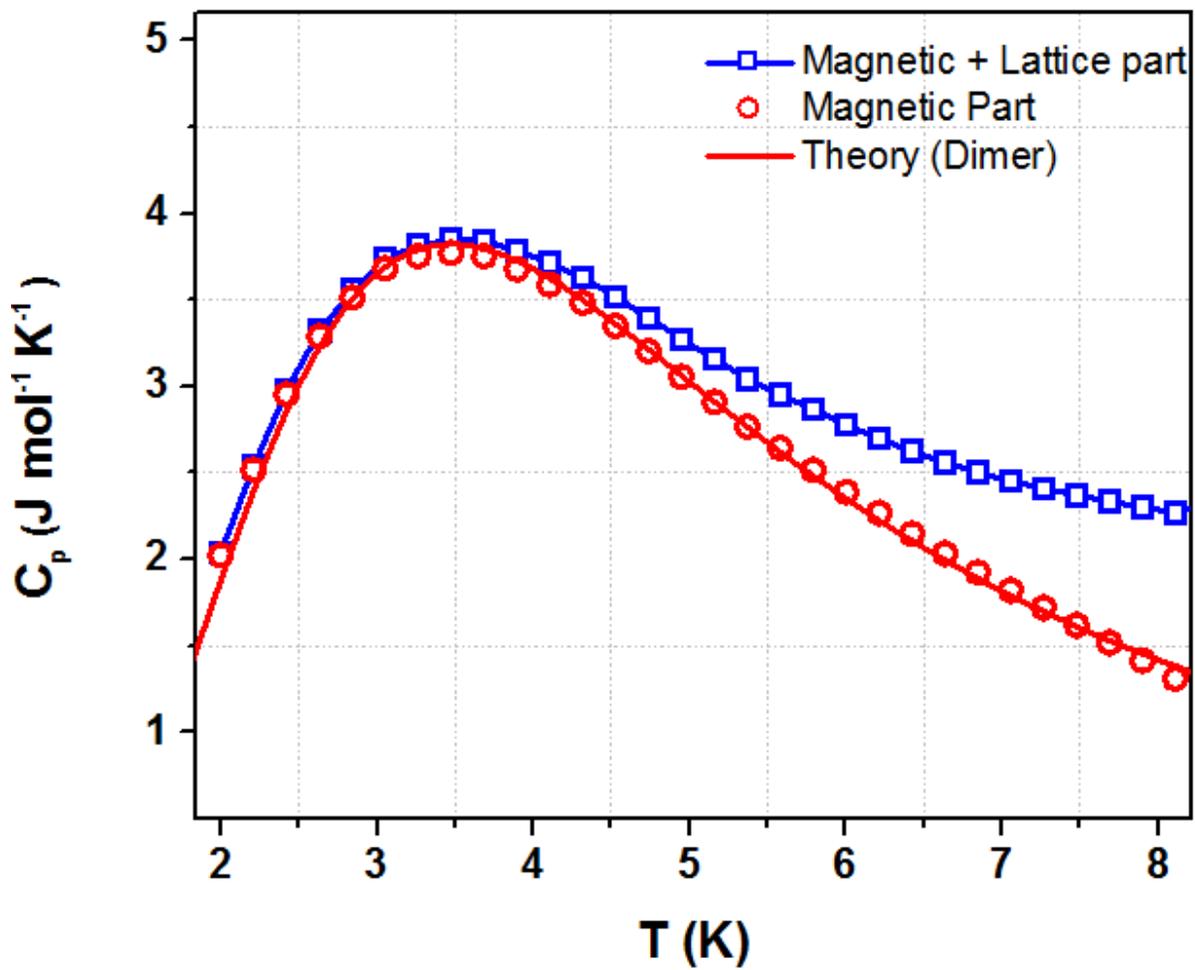

FIG. 3. Total (magnetic and lattice component) experimental specific heat is presented by the open squares. Open circles represent the magnetic contribution from the experimental data. Solid red curve shows the theoretical plot as mentioned in the text.

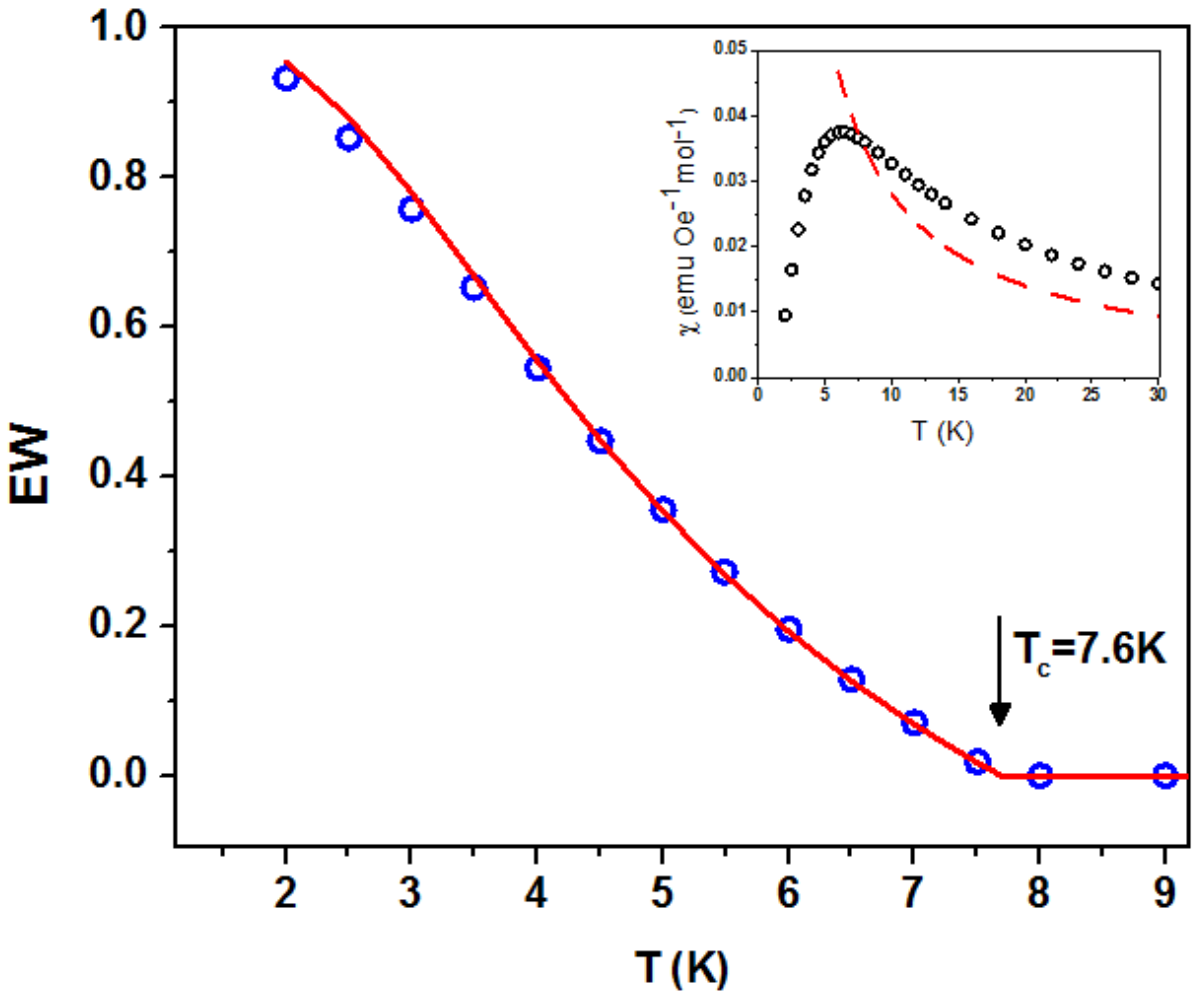

FIG. 4. Experimentally determined EW (from $\chi$) as a function of temperature. Inset shows the temperature dependent susceptibility data (circles) along with the entanglement bound (dotted red curve).

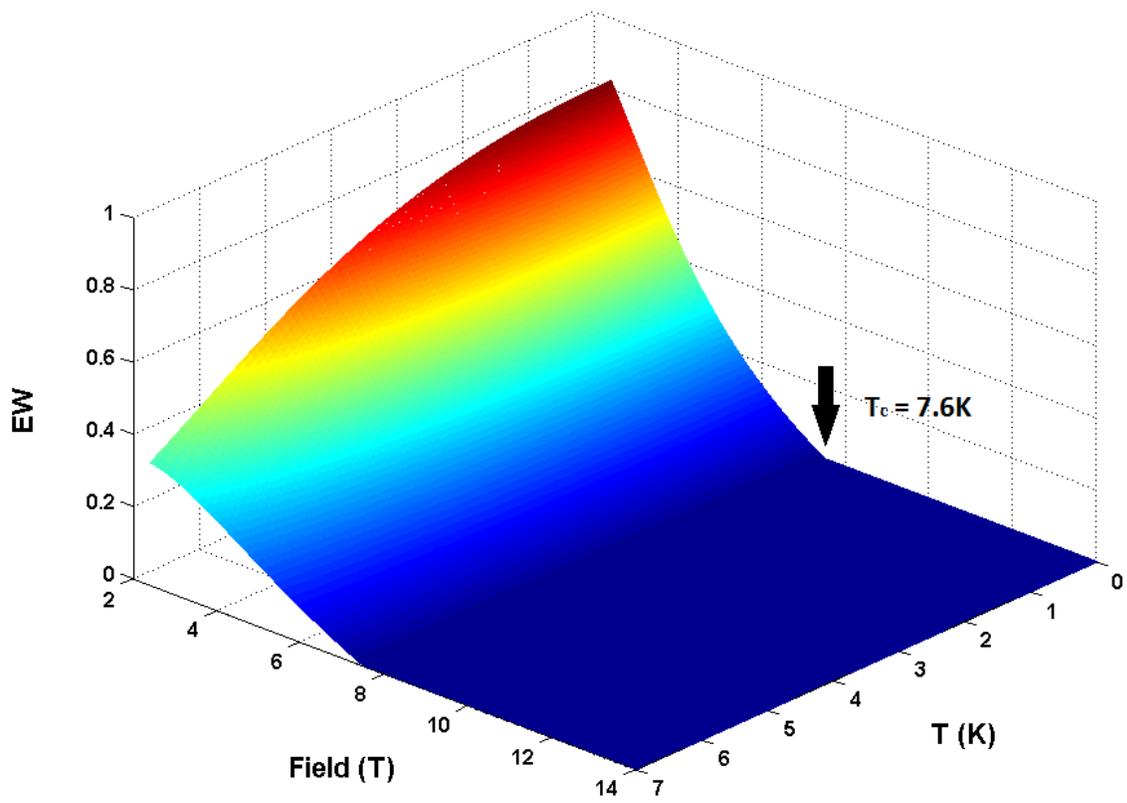

FIG. 5. Theoretically simulated EW for two qubit spin ½ Heisenberg model as a function of temperature and magnetic field.

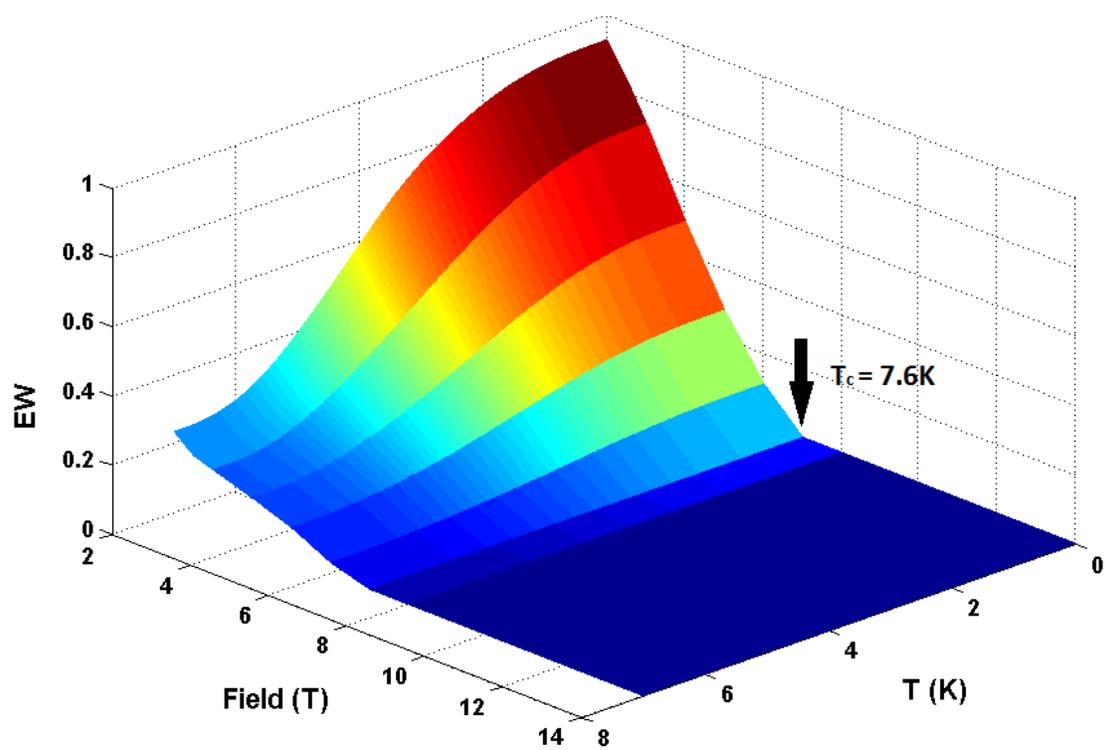

FIG. 6. Three dimensional plot showing quantified EW for $NH_4CuPO_4 \cdot H_2O$ with magnetic field and temperature along the horizontal axes.

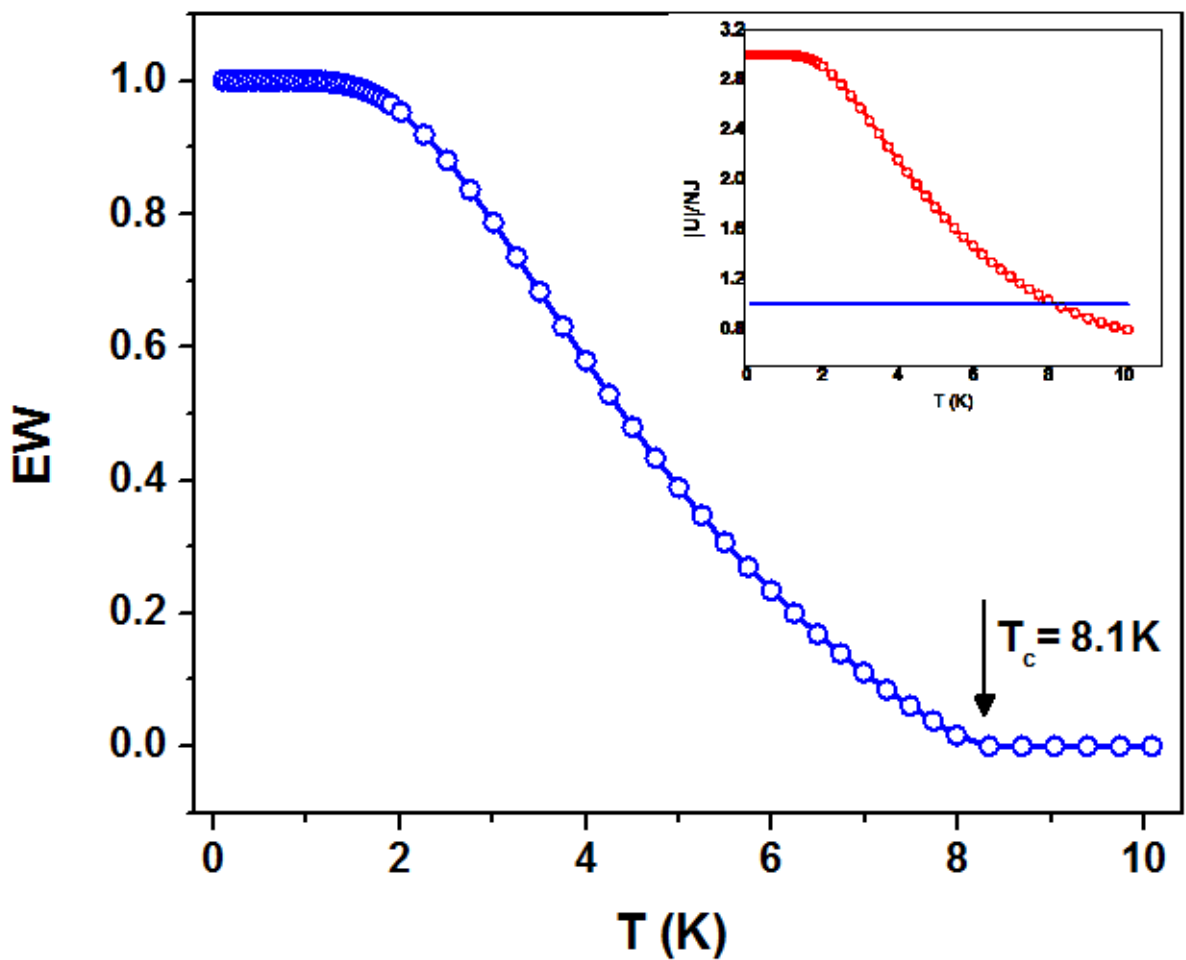

FIG. 7. Extracted entanglement from zero field specific heat data. Inset shows the plot of $\frac{2|U|}{NJ}$ (circles) with the bound (blue curve) at $\frac{2|U|}{NJ}=1$.

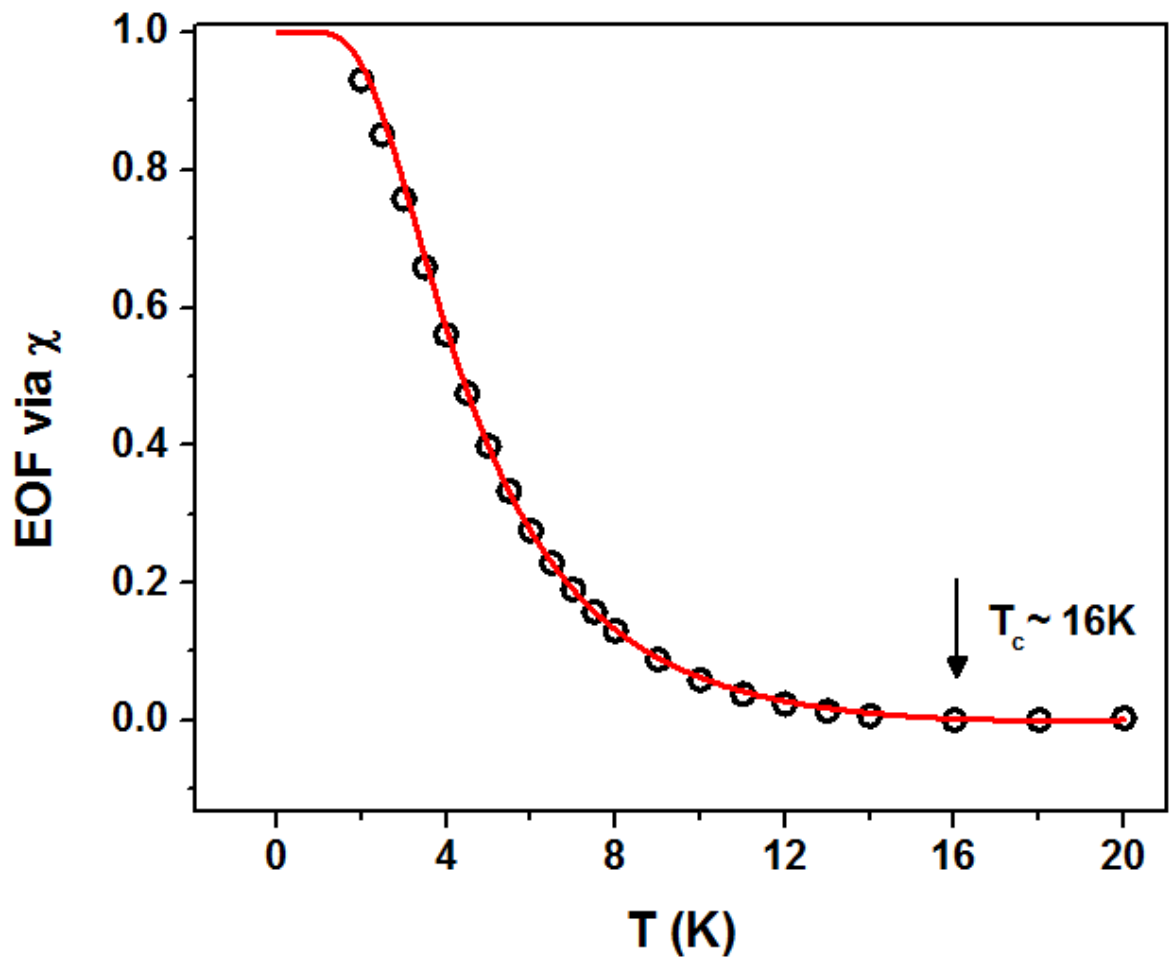

FIG. 8. Quantified EOF (circles) from magnetic susceptibility as a function of temperature. Theoretically calculated EOF is shown by the Solid red line.

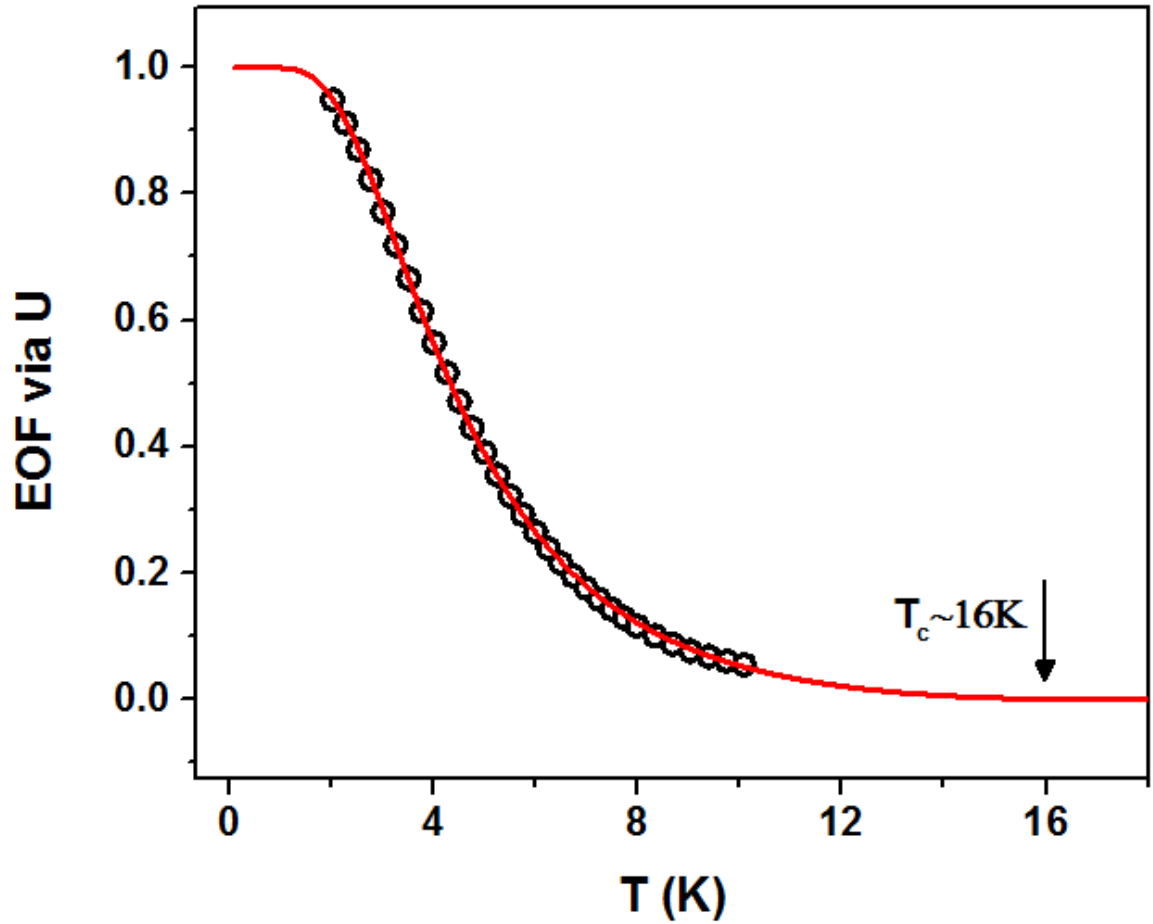

FIG. 9. Quantified EOF (circles) from internal energy as a function of temperature. Theoretical prediction for Heisenberg dimer model is shown by the Solid red line.